\documentclass{article}
\usepackage[utf8]{inputenc}
\usepackage{natbib}
\usepackage{amssymb}
\usepackage{pifont}
\usepackage{bm}
\usepackage{mathalfa} 
\usepackage{amsmath}
\usepackage[dvipsnames]{xcolor}
\usepackage{comment}
\usepackage{soul}
\usepackage{multirow}
\usepackage{listings}
\usepackage{url}
\usepackage{graphicx}

\renewcommand{\baselinestretch}{1.5}

\newcommand{\ind}{\perp\!\!\!\!\!\!\perp}

\title{Retrieved dropout imputation considering administrative study withdrawal}
\author{Rong Liu, Yongming Qu \\ Department of Global Statistical Sciences \\ Eli Lilly and Company, Indianapolis, Indiana, USA \\ }
\date{\today}

\begin{document}

\maketitle

\begin{abstract}
The International Council for Harmonisation of Technical Requirements for Pharmaceuticals for Human Use (ICH) E9 (R1) Addendum provides a framework for defining estimands in clinical trials. Treatment policy strategy is the mostly used approach to handle intercurrent events in defining estimands. Imputing missing values for potential outcomes under the treatment policy strategy has been discussed in the literature. Missing values as a result of administrative study withdrawals (such as  site closures due to business reasons, COVID-19 control measures, and geopolitical conflicts, etc.) are often imputed in the same way as other missing values occurring after intercurrent events related to safety or efficacy. Some research suggests using a hypothetical strategy to handle the treatment discontinuations due to administrative study withdrawal in defining the estimands and imputing the missing values based on completer data assuming missing at random, but this approach ignores the fact that subjects might experience other intercurrent events had they not had the administrative study withdrawal. In this article, we consider the administrative study withdrawal censors the normal real-world like intercurrent events and propose two methods for handling the corresponding missing values under the retrieved dropout imputation framework. Simulation shows the two methods perform well. We also applied the methods to actual clinical trial data evaluating an anti-diabetes treatment.  
\end{abstract}

\section{Introduction}

Estimands and missing data are crucial considerations in clinical trial design and analysis. The \cite{international2020harmonised} has established a framework for defining estimands with five key attributes: \emph{treatment} condition of interest, \emph{population}, \emph{variable}, handling of \emph{intercurrent events}, and \emph{population-level summary}. The framework includes several strategies for addressing intercurrent events, such as treatment policy, hypothetical, composite variable, while-on-treatment, and principal stratum strategies. The treatment policy strategy, which uses the potential outcome despite intercurrent events to determine the treatment effect in estimands, is gaining increasing attention in pharmaceutical statistics and has become a favored approach by regulatory agencies for many disease areas. The intent of a treatment policy strategy is to estimate the treatment effect as it would occur under real-world conditions, rather than in a hypothetical scenario where all subjects would adhere to the treatment. To achieve this, the study should implement strategies designed to systematically maximize subject retention and consistently minimize missing data, even after intercurrent events (e.g. treatment discontinuation). In cases where missing data still arise despite these precautions, the imputation of missing values will be conducted in a manner that reflects the likely real-world values that would have been observed had the data been collected. Methods to impute missing values under the treatment policy strategy have been discussed recently \citep{wang2023statistical, he2023retrieved}. These methods impute all missing values in a clinical study uniformly using one single approach, in a fashion consistent with what the values would have been had they been collected while off treatment. 

The \cite{international2020harmonised} provides clear differentiation between treatment discontinuation and study withdrawal. \cite{lipkovich2020causal} suggest using the potential outcome framework in causal inference to define estimands. In the \emph{spirit} of the treatment policy strategy, the question of interest here is that “what is the potential outcome under the treatment policy strategy if the administrative study withdrawal does not occur?” or “what is the potential outcome when using the treatment policy strategy to handle intercurrent events that similarly occur in real life and normal circumstances?”. Typically, reasons for treatment discontinuation and study withdrawal are recorded on separate case report forms (CRFs) in clinical trials. There are 2 main scenarios regarding treatment discontinuation and study withdrawal. 
Firstly, study withdrawal can occur after treatment discontinuation.
For example, in a 12-month study, a subject might stop taking the study treatment at 3 months and then
withdraw from the study entirely at 6 months. In this case, the reasons for the treatment discontinuation
and study withdrawal are generally different, such as discontinuing treatment due to an adverse event and
withdrawing from the study due to relocation. We consider treatment discontinuations in this scenario as normal real-world like (NRWL) treatment discontinuations.
Secondly, a subject might need to discontinue both the study and treatment at the same time. There are two scenarios for such cases, depending on the reason for the study withdrawal:
\begin{enumerate}
\item \textit{Administrative study withdrawal}: this occurs due to reasons beyond the participants' control (e.g.,  pandemics, geopolitical conflicts, and natural disasters) or with clearly documented reasons that are not related to treatment effect and disease progression (e.g., relocation, frequent traveling, and scheduling conflicts). In this case, continued outcome collection is not feasible, but estimating the treatment effect under such extraordinary conditions that do not reflect normal real life circumstances, or under the treatment discontinuations that would not occur in the real life (e.g., relocation generally does not prevent patients continuing the medication as they can get the medicine from the pharmacy in the new location), is not of interest. Treatment discontinuations resulting from such study withdrawals are not considered NRWL treatment discontinuations.
\item \textit{Study withdrawal possibly related to treatment or disease progression}:  This includes cases where the study withdrawal is \emph{possibly} related to study medication or disease progression, such as lost to follow-up, or the withdrawal \emph{clearly} due to reasons related to treatment or disease progression, such as adverse event or lack of efficacy. To be conservative, treatment discontinuations resulting from such study withdrawals are considered NRWL treatment discontinuations.
\end{enumerate}


\cite{darken2020attributable} and \cite{qu2021defining} consider using mixed strategies in handling different types of intercurrent events in the same study. They propose using the treatment policy strategy to handle \textcolor{brown}{NRWL} intercurrent events while using the hypothetical strategy to other intercurrent events. One drawback to this approach is that it overlooks the possibility that a subject may experience multiple intercurrent events, and these events (including study withdrawals) may have competing risks. Had this subject not experienced an intercurrent event or had not withdrawn from the study due to administrative reasons, this subject might have experienced (other) intercurrent events, such as treatment discontinuation due to adverse events or lack of efficacy. 

In this article, we propose and discuss methods for handling missing values due to study withdrawals when the treatment policy strategy is used to handle NRWL treatment discontinuations in defining estimands. Section 2 describes the statistical methods. Section 3 outlines simulation schema and presents the results of the simulation study. In Section 4, we apply this method to a clinical trial evaluating an anti-diabetes medication. Finally, Section 5 provides a summary and discussion. 
 
\section{Methods} \label{sec:methods}
Let $Y_{jk}$ denote the longitudinal outcome at time $t_k$ ($0 < t_1 < \ldots < t_K = d$) for subject $j$, where $d$ is the study duration. Let $i=0,1$ denote the control and experimental arms, respectively. Let $X_j$ be a vector of baseline covariates (including the baseline value of the response variable), and $Z_j$ is the code for the assigned treatment (0 for control and 1 for experimental treatment). Generally, $Y_{jK}$, the response at the final time point, is the primary outcome measurement. We use the notation introduced by \cite{lipkovich2020causal} for the potential outcome under the hypothetical and treatment policy strategies.  For each variable, we use ``$(i, i)$" after the variable name to denote the potential outcome under the hypothetical strategy (adhering to treatment $i$ when assigned to treatment $i$), and use ``$(i)$" to denote the potential outcome under the treatment policy strategy if assigned to treatment $i$. For example, $Y_{jk}(1,1)$ is the potential outcome under the hypothetical strategy for subject $j$ at time $t_k$ if this subject is adherent to the assigned experimental treatment, and $Y_{jk}(1)$ is the potential outcome under the treatment policy strategy if assigned to the experimental treatment. Let $U_j(i)$ denote the random variable for the time to the NRWL treatment discontinuation for subject $j$ under treatment $i$ if subject $j$ would continue the study and $V_j(i)$ denote random variable for the time to study withdrawal. Note $U_j(i)$ is censored if $U_j(i) > V_j(i)$. 
Let $D_j (i)$ denote type of study discontinuation for subject $j$ under treatment $i$: $D_j(i) = 1$ indicates study discontinuation due to administrative study withdrawal (for example site closure, geographic conflicts, pandemic) and $D_j(i)=0$ represents study discontinuation due to other reasons not clearly documented or reasons that could be related to study treatment or disease severity (for example, lost to follow-up, study noncompliance). Let $A_{jk}(i)$ denote the treatment adherence indicator at time $t_k$ for subject $j$ under treatment $i$ such that $A_{jk}(i)=1$ (adherent) if $U_j(i) \ge t_k$ and $A_{jk}(i)=0$ otherwise. Let $R_{jk}$ be the missing data indicator for $Y_{jk}(i)$ such that $R_{jk}(i) = 1$ indicates the outcome is missing and $R_{jk}(i) = 0$ means the outcome is observed. In general, $R_{jk}(i) = 1$ if $V_j(t)<t_k$ and $R_{jk}(i) = 0$ otherwise. 

In this section, we describe a procedure to estimate the treatment effect with the potential outcome consistent with the estimand using the treatment regimen policy to handle intercurrent events. 

\begin{figure} [hb!t]
\centering
\includegraphics[scale=1]{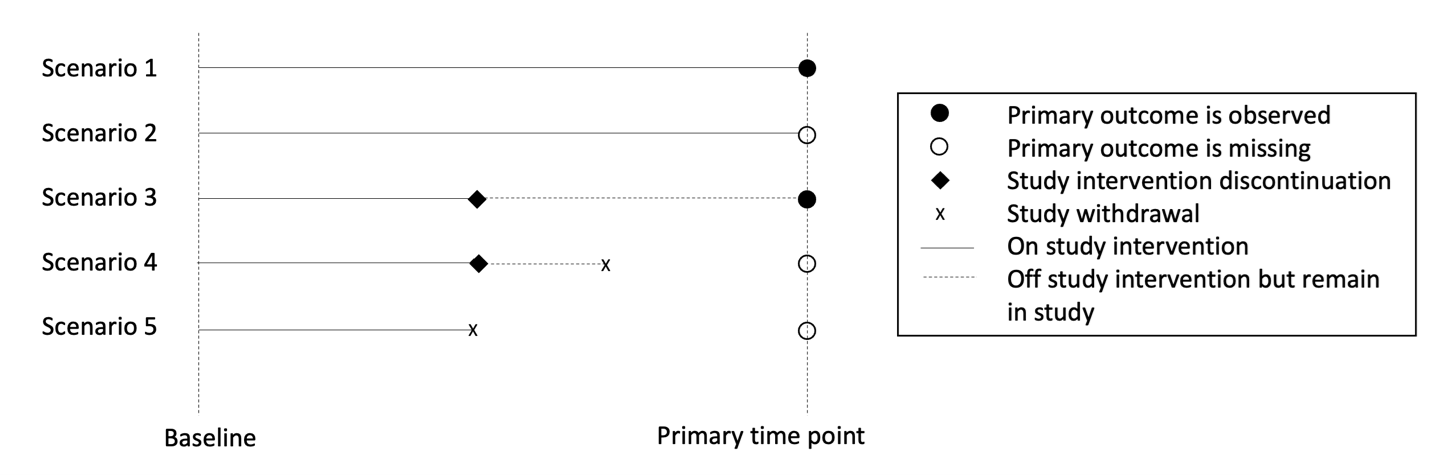}
\caption{ \small
The scenarios for missing data, the treatment discontinuation, and study withdrawal in a clinical trial of fixed duration. 
}
\label{plot:md_scenario}
\end{figure}

As illustrated in Figure \ref{plot:md_scenario}, there are 5 scenarios of missing data, the treatment discontinuation, and study withdrawal, regarding the treatment journey of a subject relative to the primary endpoint in a clinical trial with fixed duration: 
\begin{description}
    \item[Scenario 1] The subject completes the study without treatment discontinuation and without missing the primary outcome.
    \item[Scenario 2] The subject completes the study without treatment discontinuation but the primary outcome is missing. The missing values in this scenario are likely due to logistic reasons, so the missing values can be imputed using observed data across all times points from subjects that fall in Scenarios 1 and 2  under the missing at random (MAR) assumption.
    \item[Scenario 3] The subject discontinues the study treatment early but the primary outcome is collected.
    \item[Scenario 4] The subject discontinues the study treatment first, then either withdraws from or completes the study later, and the primary outcome is missing. In this case, the study withdrawal is typically assumed to be unlikely related to the potential primary outcome unless the primary outcome is correlated with subjects' ability (e.g., mobility) to continue the clinical trial. This assumption can be described in mathematical form as
    \begin{equation} \label{eq:s4}
        Y_{jK}(i) \ind V_j(i) |X_j, V_j(i) > U_j(i).
    \end{equation}
    Under this assumption, the missing values can be imputed using the retrieved dropouts, that is, subjects in Scenario 3. 
    \item[Scenario 5] The subject withdraws from the study before NRWL treatment discontinuation, leading to a missing primary outcome and censoring the time of treatment discontinuation. 
    We can handle the situation differently according to the reason for study withdrawal:
    \begin{enumerate}
        \item[5.1] For study withdrawal not due to administrative reasons, or reasons not clearly documented, we can consider that the NRWL treatment discontinuation occurs at the same time as study withdrawal, that is, 
        \begin{equation}\label{eq:s5_1}
            U_j(i)=V_j(i) |D_j(i)=0.
        \end{equation} 
        \item[5.2] For administrative study withdrawals, we consider imputing missing data in a fashion consistent with what the values would likely have been had the events leading to study withdrawl not occured and the efficacy outcomes been collected. 
        In this context, the study discontinuation time, is independent of both the potential outcome for the time to the NRWL treatment discontinuation and the potential outcome for the primary outcome.  For example, if a subject could no longer continue with the study or the treatment due to  local war, their potential time to the NRWL treatment discontinuation and potential primary outcome would not be related to their study discontinuation time due to the war. 
        \begin{equation} \label{eq:s5.2}
          V_j(i) \ind \{U_j(i) , Y_j(i)\}|X_j, D_j(i) = 1.
        \end{equation}
        In this case, we evaluated two methods to impute the missing outcomes for $Y_{jK}$, considering that some subjects that fall into this scenario could discontinue treatment due to adverse events or lack of efficacy if they were to continue in the study.         
        \begin{enumerate}
            \item[5.2(a)] Let $\hat S^{(i)}(t|X)$ denote the estimator for the survival function for $U_j(i)$ conditional on baseline covariates $X$. For each subject with an administrative study withdrawal, the estimated probability of the NRWL treatment discontinuation from time $V_j(i)$ to the study end $d$ is $$\hat p_j = P(V_j(t) < U_j(i) < d | U_j(i) > V_j(i)) = \frac{\hat S^{(i)}(V_j(t)|X) - \hat S^{(i)}(d|X)}{\hat S^{(i)}(V_j(t)|X)}.$$ 
            A binary random number $\xi_j$  can be generated from a Bernoulli distribution with mean $\hat p_j$. We assume that if $\xi_j = 1$, the subject would have discontinued treatment, and we would use retrieved dropouts to impute the missing value in $Y_{jK}(i)$. Conversely, if $\xi_j=0$, we assume that the subject would continue study treatment until the study end and use subjects who adhere to the study treatment throughout the study to impute $Y_{jK}(i)$. 
            \item[5.2(b)] Use observed and imputed values at the endpoint for subjects in Scenarios 1, 2, 3, 4, and 5.1 to impute the missing value in $Y_{jK}(i)$ for subjects in Scenario 5.2. Since the source data used for imputation includes subjects with and without NRWL treatment discontinuations, the imputed data approximately reflect that each subject in Scenario 5.2 has a certain probability of NRWL treatment discontinuations.  
        \end{enumerate}
    \end{enumerate}
\end{description}

Table \ref{table_impute_smy} provides an overview of the imputation method for each of the five scenarios.

\renewcommand{\baselinestretch}{1.0}
\begin{table}[h!tb] \centering
\caption{Imputation procedure for each scenario by treatment group}
\begin{tabular}{p{6cm} p{2cm} p{7.2cm} }
\hline\hline
Scenario (Treatment/Study Discontinuation; Missingness at Endpoint) & Assumption & Methods to Handle Missing Values in $Y_K$\\ \hline
Scenario 1. No treatment or study discontinuation; no missing value & N/A & N/A\\ & & \\
Scenario 2.	No treatment or study discontinuation; with missing value	& MAR & Use observed data from Scenarios 1 and 2 to impute under the MAR assumption.\\  & & \\
Scenario 3.	Treatment discontinuation but no study discontinuation; no missing value (retrieved dropouts) & N/A & N/A \\ & & \\
Scenario 4. Treatment discontinuation followed by study discontinuation/completion later; with missing value & Equation (\ref{eq:s4}) & Missing values at endpoint will be imputed under the assumption of multivariate normality for baseline and the endpoint visit, based upon data from Scenario 3 (retrieved dropouts). \\ & & \\
Scenario 5.	Study discontinuation resulting in treatment discontinuation; with missing value	& & \\ & & \\
\hangindent=0.5cm $\quad$ 5.1. Study withdrawal is possibly related to study treatment & Equation (\ref{eq:s5_1})  & Use data in Scenario 3 (retrieved dropouts) to impute by treatment group (refer to Scenario 4).
\\ & & \\
\hangindent=0.5cm $\quad$ 5.2: Study discontinuation is clearly due to administrative reasons & Equation (\ref{eq:s5.2}) & Method 5.2(a): Impute the NRWL treatment discontinuation status first; if imputed the NRWL treatment discontinuation status is "yes", the corresponding missing value in $Y_{jK}(i)$ is imputed using retrieved dropouts (subjects in Scenario 3); if imputed the NRWL treatment discontinuation status is "no", the corresponding missing value in $Y_{jK}(i)$ is imputed using observed data from Scenarios 1 and 2 under the MAR assumption. \\ & & \\
& &
Method 5.2(b): Use data (observed and imputed) at the endpoint in Scenarios 1 to 4, and 5.1 to impute the missing values at endpoint in Scenario 5.2.\\
\hline\hline
\end{tabular}\\
    {\begin{flushleft} Abbreviations: MAR: missing at random; N/A: not applicable. \end{flushleft} }
\label{table_impute_smy}
\end{table}
\renewcommand{\baselinestretch}{1.5}

\section{Simulations} \label{sec:simulations}


We considered a 48-week clinical trial that evaluated anti-diabetes medication. The primary endpoint was the change in Hemoglobin A1c (HbA1c) from baseline to 48 weeks, with measurements taken at baseline and then subsequently every 12 weeks throughout the study. For simplicity, we assumed that the trial only had two treatment arms: a control group and an experimental treatment group, each with a sample size of 200.  For simulating a single trial, the process involved the following 6 steps: 
\begin{description}
\item[Step 1] For each subject, we simulated the change in HbA1c from baseline across all time points under the assumption that the subject was adherent to the study treatment from an exponential decay model:
\begin{equation} \label{eq:y_po}
Y_{jk}(Z_j, Z_j) = \{\theta_{Z_j} + (\beta_0 + Z_j \beta_1) (X_{j} - \mu_x) + s_j\}\{1- e^{-\kappa t_k}\} + \epsilon_{jk},    
\end{equation}
where $j$ is the index for the subjects ($j = 1,2,...,400$); $k$ is the index for the time points ($k=1,2,3,4$), representing Week 12, 24, 36 and 48, respectively; $Z_j$ is the treatment assignment with $Z_j = 0$ for $j \le 200$ and $Z_j=1$ of $j>200$; $X_j$ is the baseline HbA1c for subject $j$; $Z_j$ is the randomized treatment (0 for control and 1 for experimental treatment) assigned to the $j$th subject; and $Y_{jk}(Z_j, Z_j)$ is the potential observation of the $j$th subject at the $k$th time point when assigned to treatment $Z_j$. We randomly drew $X_j$'s from a location-scaled Beta distribution with parameters Beta (1.5, 2), a location parameter of 7 and a scale parameter of 3, resulting in a mean baseline $\mu_x$ of approximate 8.3. In this model, $\theta_{Z_j}$ is the (theoretical) ultimate change in HbA1c for treatment $Z_j$ if we follow the subjects long enough. We chose $\beta_0$ and $\beta_1$ to be $-0.1$ and 0.2, respectively. Random terms are assumed independent and normally distributed among and within subjects: $s_j \sim N(0, \sigma_s^2)$ is the between-subject error, and $\epsilon_{jk} \sim N(0, \sigma_e^2)$ is the within-subject error. We chose $\sigma_s^2$ and $\sigma_e^2$ to be 1 and 0.5, respectively. 
\item[Step 2] We simulated the treatment adherence status $A_{jk}(Z_j)$ for subject $j$ in treatment group $Z_j$ at time point $k$ from a Bernoulli distribution with probability of 
\begin{equation} \label{eq:prob_tdc}
\Pr(A_{jk}(Z_j,Z_j) = 0 | Y_{j,k-1}(Z_j, Z_j)) = \frac{\exp(\alpha_0 + \alpha_1 Y_{j,k-1}(Z_j,Z_j))}{1 + \exp(\alpha_0 + \alpha_1 Y_{j,k-1}(Z_j,Z_j))} + c_{k}(Z_j).
\end{equation}
The first part of Equation (\ref{eq:prob_tdc}) models the dropout due to lack of efficacy, and the second part models the dropout due to adverse events or other reasons. 
If $A_{jk}(Z_j)=0$, subject $j$ discontinues treatment $i$ right after the time point $t_{k-1}$. Then, the time to the NRWL treatment discontinuation for subject $j$ is given by $t_{a,j} = \min\{t_{k-1}: A_{jk}(Z_j) = 0\}$. Through this step, we essentially simulated the time for the NRWL treatment discontinuation for all subjects. We chose $c_1(0)=c_2(0)=c_3(0)=c_4(0)=0.2$, $c_1(1)=0.06$, $c_2(1)=0.06$, $c_3(1)=0.03$, $c_4(1)=0.02$. Different values for $\alpha_0$ and $\alpha_1$ were specified for 2 simulation settings, which will be described later. 

\item[Step 3] For subjects with the NRWL treatment discontinuation occurring before the endpoint, we simulated the potential outcome under the treatment policy strategy:
\begin{equation} \label{eq:y_po_tp}
Y_{jk}(Z_j) = [\theta_{Z_j} - (\theta_{Z_j}-\theta_0)\cdot \min\{\max(t_k-t_{a,j},0), 24\}/24 + (\beta_0+Z_j\beta_1) (X_{j} - \mu_x) + s_j](1- e^{-\kappa t_k})  + \epsilon_{jk}.
\end{equation}
The above model assumes the effect of the experimental treatment on HbA1c will be washed out in 24 weeks while the effect of the control treatment will stay the same after the NRWL treatment discontinuation. It follows that
\[
Y_{jk}(Z_j) = Y_{jk}(Z_j,Z_j) + [-(\theta_{Z_j}-\theta_{0})\cdot \min\{\max(t_k-t_{a,j},0), 24\}/24] (1- e^{-\kappa t_k}).
\]
Figure \ref{plot:mean_response} illustrates the change in the mean response for various time points of NRWL treatment discontinuation.  

\begin{figure}
\centering
\includegraphics[scale = 0.6]{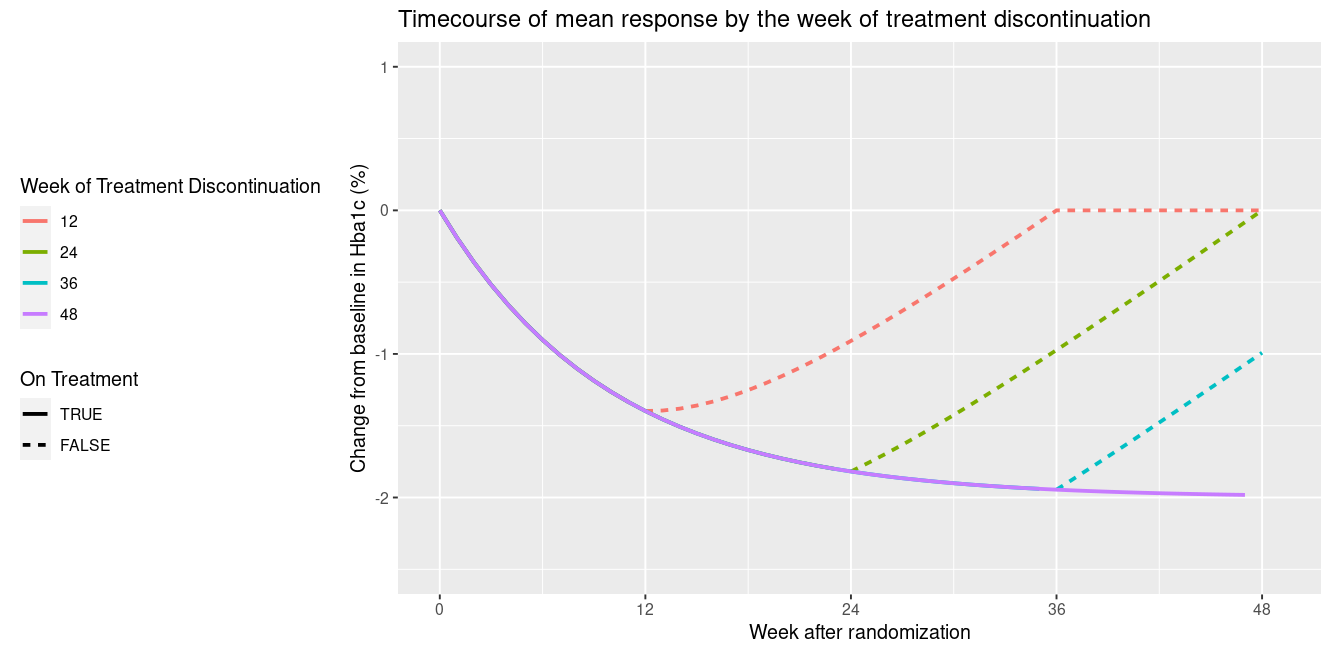}
\caption{ \small
Timecourse of mean response by various time of the NRWL treatment discontinuation
}
\label{plot:mean_response}
\end{figure}

\item[Step 4] For all subjects, simulate the time to study withdrawal due to administrative reasons from an exponential distribution assuming a constant hazard rate $\lambda$ for subjects to discontinue from the study due to administrative reasons. Through this step, we essentially simulated subjects for Scenario 5.2. Note that we only generated study discontinuations due to administrative reasons (Scenario 5.2). Given that study withdrawals due to non-administrative reasons (Scenario 5.1) are handled similarly to treatment discontinuation in Scenario 4, we can consider these discontinuations as part of the treatment discontinuations modeled in Equation (11). Subsequently, in the paper, we combined Scenarios 4 and 5.1, referring to them collectively as Scenarios 4 \& 5.1. 

\item[Step 5] A probability of missingness of 0.05 was imposed to the observations at Week 48 from subjects who had neither discontinued treatment nor withdrawal from the study. This simulated the subjects for Scenario 2. 

\item[Step 6] A probability of missingness of 0.8 was imposed to  the observations at Week 48 from subjects who had discontinued treatment before the endpoint but stayed in the study until the endpoint. This helped to control the percentage of subjects that fell into Scenario 3.  
\end{description}

\begin{table}
\caption{The average number (\%) of subjects in each scenario}
\begin{tabular}{ccccccc}
\hline\hline
{Setting}& Treatment Group & {Scenario 1} & Scenario 2 & Scenario 3 & Scenario 4 \& 5.1 & Scenario 5.2 \\ \hline
\multirow{2}{*}{1} & Control & 132 (65.8\%) & 7.0 (3.5\%) & 8.8 (4.4\%) & 36.7 (18.4\%) & 15.9 (8.0\%) \\ 
& Experimental & 136 (68.0\%) & 7.1 (3.6\%) & 7.9 (4.0\%) & 33.3 (16.7\%) & 15.7 (7.9\%) \\
\cline{2-7}
\multirow{2}{*}{2} & Control & 109 (54.7\%) & 5.8 (2.9\%) & 8.6 (4.3\%) & 40.6 (20.3\%) & 35.5 (17.8\%) \\ 
& Experimental & 125 (62.6\%) & 6.7 (3.3\%) & 5.9 (3.0\%) & 23.5 (11.7\%) & 38.7 (19.4\%) \\
\hline\hline
\end{tabular}
\label{t:simu_dist}
\end{table}

We considered 2 simulation settings, each assessed using a total of 5,000 trials simulated according to Steps 1-6 outlined above. The average number and percentage of subjects by treatment group for each setting are presented in Table \ref{t:simu_dist}.
\begin{enumerate}
    \item[] \textbf{Setting 1}: We chose $\alpha_0 = -3.5$, $\alpha_1 = 1.5$, and $\lambda = 0.002$. This setting mimics real clinical trials in evaluating anti-diabetes treatments.  
    \item[] \textbf{Setting 2}:  We chose $\alpha_0 = -3.5$, $\alpha_1 = 0$, and $\lambda = 0.005$. This setting allows for the evaluation of the performance of various methods under a scenario with a larger proportion of study withdrawals and discontinuation from treatment is purely random not depending on efficacy from previous time point(s). 
\end{enumerate}

For each simulated dataset, four estimators with different methods to handle the missing values due to study withdrawal were calculated:
\begin{enumerate}
    \item[]\textbf{Method A}: Using the non-missing values for those who were adherent to the assigned treatment to impute the missing values due to study withdrawal.
    \item[]\textbf{Method B}: Using the retrieved dropouts to impute the missing values due to study withdrawal. This method was described in \cite{wang2023statistical}.
    \item[]\textbf{Method C}: Using Method 5.2 (a) to impute the missing values due to study withdrawal.
    \item[]\textbf{Method D}: Using Method 5.2 (b) to impute the missing values due to study withdrawal.
\end{enumerate}

To evaluate the performance of these methods, the true values for the parameters of interest, including means under the treatment policy strategy for the control group, the experimental treatment group, and the treatment difference, were calculated as the average of the estimated values across 20,000 simulated completed datasets. These datasets were generated using the models described in Steps 1-3, before imposing missing data through treatment and study discontinuations on the simulated datasets.
The true means under the treatment policy strategy for the control group, the experimental treatment group, and the treatment difference are -0.001, $-1.591$, and $-1.590$, respectively in Setting 1,  and 0.001, $-1.497$, and $-1.498$, respectively in Setting 2. 

Table \ref{table_simu_results} presents the simulation results using multiple imputation to handle missing values for Methods A to D. In Setting 1, Methods A, C, and D seemed to perform well, with small bias and approximately 95\% coverage probability for the 95\% confidence interval. In Setting 2, as the proportion of administrative study withdrawal increased, Method A showed some negative bias in treatment effect, suggesting Method A is anti-conservative for estimating the treatment effect. As expected, the bias for Method B increased as the proportion of administrative withdrawal increased. Methods C and D still performed well. 

\begin{table}[hb!t]
\caption{Simulation results for the mean response for each treatment group and the treatment difference for various methods for a study with a total sample size of 400 (based on 5,000 simulated datasets)}
\begin{tabular}{p{2cm}ccccccccccc}\hline\hline
\multirow{2}{*}{Group} & \multirow{2}{*}{Method} & \multicolumn{4}{c}{Setting 1} && \multicolumn{4}{c}{Setting 2} \\
\cline{4-7} \cline{9-12}
& && {BIAS} & {ESE} & {ASE} & {CP} && {BIAS} & {ESE} & {ASE} & {CP} \\
\hline
\multirow{4}{*}{Control}
& A && -0.005 & 0.129 &0.143 &0.958&& 0.001 & 0.122 & 0.138 & 0.965\\
& B && 0.060 &0.161 &0.182 &0.942 &&  0.035 & 0.209 & 0.221 & 0.938 \\
& C && 0.012 &0.134 &0.151 &0.958 && 0.001 & 0.132  & 0.153 & 0.964\\
& D && 0.008 &0.137 &0.153 &0.957 && 0.003 & 0.140  & 0.161 & 0.962 \\
&&&&&&&&&&& \\
\multirow{4}{*}{Treatment} 
& A && -0.012 &0.141 & 0.142 &0.931 && -0.058 & 0.153 & 0.158 & 0.907\\
& B && 0.143 &0.177 &0.190 &0.857 && 0.229 & 0.236 &	0.232 & 0.781\\
& C && 0.004 &0.149 &0.157 &0.933 && -0.012& 0.163 &	0.164 & 0.923\\
& D && 0.022 &0.151 &0.154 &0.931 && 0.022 & 0.181 &	0.188 & 0.932\\ 
&&&&&&&&&&& \\
\multirow{4}{2cm}{Treatment Difference} 
& A && -0.007 &0.191 &0.193 &0.939 && -0.058 & 0.197 & 0.200 & 0.927 \\
& B && 0.084 &0.240 &0.245 &0.932  && 0.195  & 0.322 & 0.292 & 0.862 \\
& C && -0.008 &0.200 &0.209 &0.947 && -0.013 & 0.212 & 0.213 & 0.934 \\
& D && 0.014 &0.203 &0.206 &0.941  && 0.019  & 0.231 & 0.232 & 0.936 \\
\hline\hline
\end{tabular}
  {\begin{flushleft} Note:  BIAS, empirical bias; CP,  95\% empirical coverage probability; ASE: mean standard error estimates of the mean based upon Rubin's rule (Rubin 1987); ESE: standard deviation of the estimates. \\ 
  \end{flushleft}}
\label{t:simdir}
\label{table_simu_results}
\end{table}

\section{Application}
We applied the four methods to estimate the change in HbA1c from baseline to 52 weeks in IMAGINE-5 Study 
(ClinicalTrials.gov Identifier: NCT01582451). IMAGINE-5 is a randomized study to compare the efficacy and safety of insulin peglispro with insulin glargine in subjects with type 2 diabetes who were previously treated with a basal insulin. There were 466 subjects randomly assigned to insulin glargine or insulin peglispro with a 1:2 randomization. The study was conducted in accordance with the Declaration of Helsinki guidelines on good clinical practices \citep{world2010world}. The primary results were published in \cite{buse2016randomized}. 

Table \ref{t:app_ice} presents the the number and percent of subjects in each scenario (defined in Table \ref{table_impute_smy}) by treatment group.
Table \ref{t:app_results} summarizes the results for the estimated mean change from baseline to 52 weeks for each treatment group along with the treatment differences for Methods A, B, C, and D. The results based on the four methods were very similar. This is likely because (1) the proportion of administrative study withdrawal is small, and (2) the mean response in the change in HbA1c from baseline to 52 weeks between retrieved dropouts and adherers was relatively small. The observed mean changes in HbA1c for adherers were $-0.236$ and $-0.672$ for insulin glargine and insulin peglispro, respectively. The observed mean changes in HbA1c for retrieved dropouts were $-0.060$ and $-0.031$ for insulin glargine and insulin peglispro, respectively.  

\begin{table}[hbt]
\caption{Summary of Number (\%) of subjects in each scenario (IMAGINE-5)}
\begin{tabular}{cccccc}
\hline\hline
{Arm}&{Scenario 1}&{Scenario 2} & {Scenario 3} & {Scenarios 4/5.1} & {Scenario 5.2}\\
\hline
{Insulin Glargine (N=159)}&{130 (81.8\%)}&{1 (0.6\%)}&{5 (3.1\%)}&{7 (4.4\%)}&{9 (5.7\%)}\\
{Insulin Peglispro (N= 307)}&{257 (83.7\%)}&{0 (0.0\%)}&{13 (4.2\%)}&{19 (6.2\%)}&{13 (4.2\%)}\\
\hline\hline
\end{tabular}
\label{t:app_ice}
\end{table}

\begin{table}
\caption{Summary of the estimated mean change in HbA1c from baseline to 52 weeks (IMAGINE-5)}
\begin{tabular}{cccc}
\hline\hline
{}&{Insulin Glargine}&{Insulin Peglispro} & {Treatment Difference}\\
{Method}&{Mean (SE)}&{Mean (SE)}&{Mean (95\% CI)}\\
\hline
A & -0.230 (0.061) & -0.602 (0.045) & -0.372 (-0.519,-0.224) \\
B & -0.219 (0.061) & -0.588 (0.045) & -0.368 (-0.517, -0.219) \\
C & -0.233 (0.061) & -0.603 (0.044) & -0.369 (-0.517, -0.221) \\
D & -0.228 (0.061) & -0.608 (0.045) & -0.380 (-0.528, -0.232) \\
\hline\hline
\end{tabular}
\label{t:app_results}
\end{table}

\section{Summary and Discussion}
Since the release of ICH E9 (R1), treatment policy strategy has become one of the most popular approaches to handle intercurrent events including premature treatment discontinuations in defining estimands. The rationale for using data after treatment discontinuation in the analyses is that it reflects real clinical use. However, treatment discontinuation caused by study withdrawal due to administrative reasons such as COVID-19 control measures, geopolitical conflicts, and site closures due to business reasons which are rare events in real life and are unlikely the interest of the estimation. Relocation, frequent traveling, and scheduling conflicts, which caused study withdrawal and treatment discontinuation, unlikely prevent patients from continuing the medication in real life. Therefore, treatment effect under thess types of treatment discontinuations that only present in clinical trials is unlikely the clinical interest. Therefore, a hypothetical strategy may be used to handle these non-NWRL treatment discontinuations. 

Hybrid estimands have recently been proposed to use mixed strategies to handle different types of intercurrent events in the same study. For example, the treatment policy strategy is used to handle treatment discontinuations due to adverse events and a hypothetical strategy is used to handle treatment discontinuations as a result of study withdrawal due to administrative reasons. However, the hybrid estimand ignores the fact that treatment discontinuations due to different reasons are at competing risks. Subjects may experience treatment discontinuations due to reasons related to efficacy or safety should they have not had the administrative study withdrawal. In this research, we propose a new method to estimate the estimand with NRWL intercurrent events handled by the treatment policy strategy assuming the administrative study withdrawal censors the NRWL treatment discontinuations. 
For participants with administrative study withdrawal, the potential status of NRWL intercurrent events censored by the study withdrawal is first imputed. Then, the missing outcome is imputed based on the imputed status of the NRWL intercurrent events: if the imputed intercurrent event status is ``no", we impute the missing value for the outcome using MAR; if the imputed intercurrent event status is ``yes", we compute the missing value using retrieved dropouts. An alternative and simpler approach is to impute the missing values due to administrative study withdrawal using all subjects with observed values (including those who are adherent and the retrieved dropouts), which naturally implies a certain proportion of subjects with administrative study withdrawal have NRWL treatment discontinuations should they have not withdrawn from the study.
These approaches make the imputation methods consistent with the potential outcome of interest under the treatment policy strategy for the NRWL intercurrent events. 

Simulation shows when the administrative treatment discontinuation rate is low [e.g., less than 10\% in Setting 1 of the simulation (Table \ref{t:simu_dist})], imputing the missing data using adherers (Method A) produces little bias (Table \ref{table_simu_results}). This is likely because only a small proportion of the administrative withdrawals would have NWRL treatment discontinuations had they not withdrawn from the study. Therefore, Method A, which is much simpler, may be used if the proportion of the administrative study withdrawals is less than 10\%.

Note administrative study withdrawals due to frequent traveling, and scheduling conflicts, which are often reported by patients, are more subjective than those caused by site closures due to pandemics, natural disasters, and geopolitical conflicts. In application of this method, the key stakeholders including sponsors and regulators should get aligned on the definition of estimands, especially on what the intercurrent events are handled by the treatment policy strategy.

To effectively use the proposed imputation method for estimands using the treatment policy strategy to handle relevant intercurrent events, it is critical to collect the accurate reasons for treatment discontinuations and study withdrawals. \cite{qu2022accurate} reviewed a set of clinical trials for the reasons of treatment/study discontinuations and suggested ways to improve the collection of the reasons for treatment/study discontinuations. An ongoing PHUSE working group on ``Implementation of Estimands (ICH E9 (R1)) using Data Standards" is developing a new recommendation for the CRF for treatment and study discontinuations \citep{PHUSE2024}. 

The competing risk for different types of intercurrent events can also be applied to situations where a hypothetical strategy is used for certain types of intercurrent events. For example, if in the estimand the intercurrent event of using rescue medication is handled by a hypothetical strategy, the data collected after the initiation of the rescue medication may be discarded. Instead of imputing the missing outcome based on the MAR assumption, we may need to treat the rescue medication use as an event censoring the NRWL treatment discontinuation. However, the probability of treatment discontinuation would likely be high had rescue not been administered. Note that, in this case, it may not be appropriate to classifying the event of using rescue medication as Scenario 5.2 and use the method proposed in this article to impute the missing (or unobserved) values. Further research is needed on this topic.

In the estimation of the cumulative distribution function for the NRWL treatment discontinuation, the proposed methods assume that the the occurrence of the NRWL treatment discontinuation is independent of the potential outcome for the analysis variable. This assumption may not hold. For example, subjects with a poor response at early time points may have a higher probability of discontinuing treatment. However, modeling the NRWL treatment discontinuations conditional on the intermediate outcomes may be complex. Sensitivity analyses are generally recommended to investigate the impact of underlying assumptions on missing data.  

On a practical note, ``same time" for Scenario 5 does not imply the exact same calendar date for clinical visits. A subject may decide to discontinue treatment and withdrawal from the study due to administrative reasons at the same time, but the formal discontinuation from the study may require a follow-up visit (which generally occurs later) to complete the discontinuation procedures. In practice, Scenario 5 can be identified by the absence of office visits between the discontinuation of treatment and the study withdrawal.

\section*{Acknowledgement}
We would like to thank Dr. Jianghao Li (Eli Lilly and Company) for reviewing the manuscript, the simulation and data analysis program and for his valuable feedback. Additionally, we extend our gratitude to Dr. Yanyao Yi (Eli Lilly and Company) for some discussions regarding this topic.  

\section*{Disclosure statement}
Both authors are employees and minor shareholders of Eli Lilly and Company. Not additional funds were received for this research.

\newpage
\bibliographystyle{apalike}
\bibliography{references}
\end{document}